\documentstyle[11pt,newpasp,twoside]{article} 
\def\edcomment#1{\iffalse\marginpar{\raggedright\sl#1\/}\else\relax\fi} 
\reversemarginpar 
 
\begin{document} 
\title{Radio Pulsar Death} 
 \author{Bing Zhang} 
\affil{The Pennsylvania State University, USA} 
 
\begin{abstract} 
Pulsar radio emission is believed to be originated from the
electron-positron pairs streaming out from the polar cap region.
Pair formation, an essential condition for pulsar radio emission, is
believed to be sustained in active pulsars via one photon process
from either the curvature radiation (CR) or the inverse Compton
scattering (ICS) seed photons, or sometimes via two photon
process. In pulsars with super-critical magnetic fields, some more 
exotic processes, such as magnetic photon splitting and bound pair
formation, will also play noticeable roles. All these effects should
be synthesized to discuss radio pulsar death both in the conventional
long-period regime due to the turnoff of the active gap, and
in the high magnetic field regime due to the possible suppression of
the free pair formation. Here I briefly review some recent progress in
understanding radio pulsar death.
\end{abstract} 
 
\section{Why pair production essential?} 

A pulsar drags its magnetosphere to co-rotate. In the observer's rest
frame, the local charge density required for co-rotation is the 
``Goldreich-Julian'' (1969) density, which when $r \ll r_{lc}$ ($r_{lc}$
is the light cylinder radius) is approximately $\rho_{_{\rm GJ}}
\simeq -({\bf \Omega \cdot B})/({2 \pi c})$, where ${\bf \Omega}$ is
the angular velocity of the star, and ${\bf B}$ is the local magnetic
field vector. In a simplest GJ magnetosphere, the positive and
negative charges are separated from each other in space, and pair
production is not required.

Pair production comes in for two reasons. First, it is obligatory. The 
most prominent feature of pulsar radio emission is its very high
brightness temperature (typical value $T_B \sim 10^{25}-10^{30}$ K).
Due to self-absorption, the maximum brightness temperature for any
incoherent emission is limited by the kinetic energy of the emitting
electrons, i.e. $T_{incoh,max} = \gamma_e m_e c^2/k \sim 6\times 10^{12}
\gamma_{e,3}$ K. The pulsar radio emission mechanism therefore must be
{\em extremely} coherent. Guided by our understanding of other
coherent sources in the universe, the emission source is very likely a
plasma in which various plasma instabilities can be developed to
achieve coherent emission. A pair plasma is therefore required in an
otherwise charge-separated magnetosphere. 

Second, it is inevitable. A rotating magnetic pulsar is a unipolar
generator, with a potential drop across the open field line region
\begin{equation}
\Phi=(B_p R^3 \Omega^2/2c^2) = 6.6\times 10^{12}~{\rm V}~B_{p,12}
P^{-2} R_6^3, 
\end{equation}
where $B_p$ is the surface magnetic field at the pole, $P$ is the
rotation period, $R$ is the neutron star radius, and the
convention $Q=10^n Q_n$ has been adopted. Such a huge potential is
likely dropped along the open field lines (see \S2 for the
reasons), which accelerates a test particle up to an energy of
$\gamma_p m_e c^2 = e \Phi \sim 6.6$ TeV, or $\gamma_p
\sim 10^7$ ($\gamma_p$ is the Lorentz factor of the ``primary''
particles to be distinguished with the secondary pairs). These
particles emit curvature radiation (CR) or inverse Compton scattering
(ICS) photons (eqs.[2-4]) during
acceleration. These primary $\gamma$-rays inevitably materialize in
a strong magnetic field or a hot thermal photon bath near the surface
(see \S2 for detailed discussions).
 
\section{Pulsar inner gaps and pair production mechanisms}

In a force-free magnetosphere where $\rho=\rho_{_{\rm GJ}}$, the
electric field parallel to the local magnetic field, $E_\parallel$, is
equal to zero everywhere. In order to have particles accelerated in a
pulsar magnetosphere, there must be a deficit of $\rho$ with respect
to $\rho_{_{\rm GJ}}$ so that an $E_\parallel$ can be developed. There
are two preferred charge-depleted regions, namely gaps, in a
pulsar magnetosphere: the polar cap region above the neutron
star surface (inner gap), and the region near or beyond the ${\bf
\Omega \cdot B}=0$ ``null charge surface'' (outer gap). Since outer
gaps can only be developed in young and millisecond pulsars while a
much larger population of pulsars have radio emission, it is generally
believed that radio emission may not be closely related to the
existence of the outer gap. Rather, it should be closely related to an
active inner gap which provides copious pairs from the polar cap
cascade.

There are two ways to classify the inner gaps. The first way is
according to the boundary condition at the pulsar surface, and
inner gaps can be divided into two subtypes: a vacuum gap 
($E_\parallel \neq 0$ at the surface, Ruderman \& Sutherland 1975) and
a space-charge-limited flow ($E_\parallel=0$ at the surface, Arons \&
Scharlemann 1979). The first type (hereafter VG) requires that the
charges (usually positive ions) are tightly bound in the surface, so
that a large vacuum gap naturally develops as charges in the polar
regions move away from the surface due to centrifugal forces. Such a
suggestion, however, is questioned later since calculations show that
ion binding energies are usually much smaller than the one required
for binding (e.g. Usov \& Melrose 1995 for a review). The picture may
be still pertained by conjecturing either that some pulsars are more
exotic objects with exposed strange quark surface (Xu, Qiao \& Zhang
1999; Xu, Zhang \& Qiao 2001), or that much stronger (compared with
the dipolar field inferred from the $P$ and $\dot P$ data) sun-spot
like local magnetic fields anchor in the polar caps (Gil \& Mitra
2001; Gil \& Melikidze 2002). The second type of inner gaps (hereafter
SCLF) is the natural outcome of a neutron star (or a strange star with
a normal matter crust) with none or weakly distorted dipolar magnetic
field configuration. In such a case, charges (either electrons for
${\bf \Omega \cdot B} >0$, or ions for ${\bf \Omega \cdot B} <0$) will
be freely pulled out from the surface. Taking $\rho =\rho_{_{\rm GJ}}$
at the surface, it is natural that a deficit of $\rho$ with respect to
$\rho_{_{\rm GJ}}$ will grow as the current flow outwards in the open
field line region, since both $\rho$ and $\rho_{_{\rm GJ}}$ follow
different $r-$dependences. In the early years of pulsar theories
when general relativistic effects are not noticed, such $\rho$-deficit
growth was believed to be due to the curvature of the pulsar dipolar
field lines (Arons \& Scharlemann 1979; Arons 1983). In an oblique
pulsar, for those field lines curving towards the rotational axis (the
so-called favorably curved lines), an decreasing angle between ${\bf
\Omega}$ and ${\bf B}$ result in a net gain in $\rho_{_{\rm GJ}}
\propto {\bf \Omega\cdot B}$ with respect to the local $\rho \propto
|{\bf B}| \propto r^{-3}$ ($r$ is the radial coordinate of the point
of interest) value defined by the magnetic flux conservation. The
small $E_\parallel$ gradient generated from a small $(\rho-\rho_{_{\rm
GJ}})$ deviation results in an elongated gap compared with VGs. The
growth of $E_\parallel$ at the edges of the gap is even smaller,
resulting in a ``slot gap'' (Arons 1983). Later Muslimov \& Tsygan
(1992) first realized that by taking into account the general
relativistic effect, another more prominent acceleration mechanism in
the SCLF picture is available. For a Kerr metric, an observer at
infinity views that local inertia frames are dragged by the rotating
body. This modified the requires ``co-rotating'' magnetosphere charge
density to a relatively smaller value: $\rho_{_{\rm GJ}}({\rm GR})
\simeq -[({\bf \Omega-\Omega_{\rm LIF}) \cdot B}] / 2\pi c \alpha
\simeq -({\bf \Omega \cdot B}/2\pi c\alpha) [1-\kappa_g (R/r)^3]$,
where $\alpha \sim 0.78$ and $\kappa_g
\sim 0.15-0.27$ are some constants depending on the equation of state
of the neutron star (Muslimov \& Harding 1997; Harding \& Muslimov
1998, hereafter HM98). Clearly, besides the ``flaring'' term due to
field line curvature (the $[{\bf \Omega \cdot B}]$ factor), an
additional $r$-dependence of $\rho_{_{\rm GJ}}$ is introduced (the
$[1-\kappa_g (R/r)^3]$ factor). It turns out that this component
results in a larger $(\rho-\rho_{_{\rm GJ}})$ deviation, and a much
faster growth of $E_\parallel$, as long as the inclination angle is
not very close to 90$^{\rm o}$. Such an effect significantly
influences the properties of the SCLF gaps (HM98). In the VG scenario,
however, since $\rho = 0$ in the gap, the frame-dragging modification
is only minor, i.e., by a factor of $(1-\kappa_g) \sim 0.85$ (Zhang,
Harding \& Muslimov 2000, hereafter ZHM00). 

The second way to classify inner gaps is according to the source of
$\gamma$-ray photons that trigger the pair cascade. For a long time
since the early pulsar theories, curvature radiation (CR) is regarded
as the only source of the primary $\gamma$-rays (Sturrock 1971;
Ruderman \& Sutherland 1975; Arons \& Scharlemann 1979). The typical
CR photon energy is (${\cal R}$ is the field line curvature)
\begin{equation}
\epsilon_{_{\rm CR}}=(3/2) (\hbar c/{\cal R}) \gamma_p^3 \simeq 76 ~{\rm
GeV}~ {{\cal R}_6}^{-1} \gamma_{p,7}^3~.
\end{equation}
The importance of inverse Compton scattering (ICS) off the thermal
X-rays near the neutron star surface is gradually appreciated later
(e.g. Xia et al. 1985; Daugherty \& Harding 1989; Dermer 1990; Sturner
et al. 1995). Zhang \& Qiao (1996) and Luo (1996), within the
frameworks of VG model and the 
(non-relativistic) SCLF model, respectively, first clearly suggested
that the ICS photons may be more energetic and therefore have shorter
mean free path to generate pairs than the CR photons. Zhang et al. 
(1997) further pointed out that there are two preferred ICS photon
energies (though K-N limit ignored), so that by combining the CR
typical energy, there are altogether three inner gap ``modes''. 
The understanding of these three modes were greatly advanced by HM98,
Hibschman \& Arons (2001a, 
b, hereafter HA01a, HA01b) and Harding \& Muslimov (2001, 2002,
hereafter HM01, HM02) within the framework of the relativistic SCLF
models. The first ICS mode is defined by the 
characteristic photon energy due to ``resonant'' scatterings (which
are generally related to transition of 
electrons between the first Landau state and the ground state, and
have much larger cross section than the Thomson cross section), i.e.,
($B'=B/B_q$ and $B_q \equiv m_e^2 c^3/\hbar e \simeq 4.414 \times
10^{13}$ G is the critical field) 
\begin{equation}
\epsilon_{_{\rm ICS,R}}=2 \gamma_p B' (m_e c^2)
\simeq 23 ~{\rm GeV}~ \gamma_{p,6} B_{12}.
\end{equation}
Scatterings above the resonance (non-resonant scatterings) also
contribute to a significant higher energy component in the final IC
spectrum. The typical energy of these scatterings is defined by the
minimum of the IC-boosted thermal peak energy and the Klein-Nishina
limit (i.e. the electron's kinetic energy)
\begin{equation}
\epsilon_{_{\rm ICS,NR}}={\rm min} (\gamma_p^2 k T, \gamma_p m_ec^2)
={\rm min} (8.6 \gamma_{p,4}^2 T_6, 5.1\gamma_{p,4})~{\rm GeV}~. 
\end{equation}
The K-N limit takes over as long as $\gamma_p > 5.9\times 10^3
T_6^{-1}$. Resonant scattering also reaches K-N regime when $B'>1$.

Gamma-rays produced via CR or ICS can materialize essentially in
two ways, i.e., one photon production (1p: $\gamma \rightarrow
e^{+}e^{-}$) in strong magnetic fields and sometimes two photon
production (2p: $\gamma\gamma \rightarrow e^{+}e^{-}$). The 1p is
believed to be the dominant source of pairs (Sturrock 1971). The pair
generation rate exponentially grows as 
the factor $(1/2)(\epsilon_\gamma / m_ec^2) B' \sin \theta_{\gamma B}$ 
reaches a critical value of $\sim 1/10$, or in high magnetic fields
($B'>0.1$), as the threshold condition $(1/2)(\epsilon_\gamma /
m_ec^2) \sin \theta_{\gamma B} > 1$ is reached (Daugherty \& Harding
1983), where $\theta_{\gamma B}$ is the incident angle between the
gamma-ray and the local magnetic field. These define the {\em ``mean
free path'' (i.e. attenuation length) of the photons}, $l_{ph}$, which
is shorter for a larger 
$\epsilon_\gamma$, a larger $\theta_{\gamma B}$ (or a shorter pulsar
period which gives a larger field line curvature), and a higher $B$
when the near-threshold effect is not important. In most pulsars,
usually more than one generation of pairs are produced. The first
generation pairs lose their perpendicular energies through synchrotron 
radiation (SR), and the resultant secondary $\gamma$-rays also usually
meet one photon pair production condition, so that a photon-pair
cascade develops (Daugherty \& Harding 1982, 1996). The remaining
parallel energy of the pairs will also be converted to radiation
through resonant ICS, and the resulting photons may sometimes
(although the condition is more stringent) be converted to further
pairs, leading to a full polar cap cascade (Zhang \& Harding 2000a). 

The 2p process is in principle expected, since the typical energy of
primary gamma-rays, $\epsilon_\gamma$, and the typical energy of the
thermal X-rays, $\epsilon_x$, usually satisfy the kinetic condition
$\epsilon_\gamma \epsilon_x (1-\cos \theta_{\gamma x}) \geq (m_e
c^2)^2$, where $\theta_{\gamma x}$ is the incident angle between the
gamma-ray and the X-ray (Zhang \& Qiao 1998). However, the mean free
path of a typical gamma-ray for 1p is usually smaller than that of 2p, 
so that a gamma-ray usually has materialized in the strong magnetic
field before interacting with the thermal X-rays. To have 2p competitive
against 1p, one has to either raise the thermal luminosity and
$\theta_{\gamma x}$ or considerably lower the magnetic field. The hot
magnetar environment (Zhang 2001) and the low-$B$ millisecond pulsars
(Harding, Muslimov \& Zhang 2002, hereafter HMZ02) are therefore
possible sites where the 2p process may become noticeable.

\section{Energy budget deathlines and death valley}

We have shown above that pair production is generally expected in
normal pulsars, and in \S1 we have made the argument that pair
production is almost obligatory in the current radio emission models. 
{\em Production of free pairs is the essential condition for pulsar
radio emission. The condition that free pair production is prohibited
or suppressed therefore defines radio pulsar death} (recent
discussions on this topic include, e.g. Arons 2000; ZHM00; HA01a,b;
HM02; HMZ02). One important caveat is that free pair production may
not be {\em the sufficient condition} for pulsar radio emission, which
is model-dependent and 
poorly known. There exists a missing link between the pulsar high
energy emission theories from which the cascade pair properties are
predicted, and the pulsar radio emission theories in which various
pair properties are required (but see Arendt \& Eilek 2002 for a
recent attempt to connect the missing link). 

When a radio pulsar ages, it slows down so that the unipolar potential 
drops. The accelerated particles achieve less energies, and their
subsequent CR/ICS photons are less energetic. Eventually, these
photons no longer pair produce, either via 1p or 2p. No further pair
plasma is ejected into the plasma, and the pulsar stops shining in
the radio band. A radio pulsar dies. The death of this type is due to
inadequate rotational energy budget. The condition defines the so-called 
energy budget ``deathline'' (e.g. Ruderman \& Sutherland 1975 and
many literatures thereafter) in the $P-B_p$ space or $P-\dot P$ space.
For a star-centered dipole, the surface magnetic field at the pole
reads ($I$ and $R$ are moment of inertia and radius of the neutron
star, respectively) 
\begin{equation}
B_p=6.4\times 10^{19}~{\rm G}~(P \dot P)^{1/2} I_{45}^{1/2} R_6^{-3}~. 
\end{equation}

In some work where detailed theories are not needed, the pulsar
deathline is conventionally taken as a line of constant $\Phi$
(eq.[1]), with a slope 3 in the $P-\dot P$ diagram. We'd like to
caution here, however, that the real death lines could considerably
differ from such a line, since the final potential drop across the gap
at the pulsar death is defined by the condition of pair production,
which gives a lower $\Phi$ with a different slope than 3.
In fact, it is not easy to draw a single line to
define radio pulsar death, since many factors will affect the position 
of the deathline.
(This is why I did not present any analytical expression of the
deathline in this chapter.) 
What is more relevant should be a ``death valley'', a term first
invented by Chen \& Ruderman (1993). 

Before moving into detailed discussions about the energy budget
deathlines, two more length parameters need to be introduced. One
is {\em the acceleration length}, $l_{acc}$, which is the typical
distance an electron has to travel for acceleration to achieve a
typical Lorentz factor $\gamma_{p,c}$, with which the electron's
CR/ICR emission photons can pair produce. This is
acceleration-model-dependent (VG vs. SCLF). Another 
one is {\em the mean free path of the electrons}, $l_e$, which is the
distance of the electron travels before emitting one CR or ICS
photon, i.e., $l_e \simeq (\epsilon_\gamma / \dot \gamma m_e c^2)
c$. This is radiation-mechanism-dependent (CR 
vs. ICS). These two length parameters, together with {\em the mean
free path of the photon}, $l_{ph}$, as discussed in \S2, are essential
for the following discussions. In all the models, a test particle has
to be accelerated through a length scale of $l_{acc}$ to reach a
Lorentz factor $\gamma_c$. This particle emits a test photon (via CR
or ICS) with a probability (depends on $l_e$), and the photon
attenuation length is $l_{ph}$. By minimizing the length $l_{acc}+ 
l_{ph}$, one gets a typical height of the gap, and hence, the
potential drop across the gap ($\Delta V$). By demanding $\Delta V$
equal to the maximum achievable potential (some fraction
of $\Phi$ in eq.[1]), one gets a deathline (ZHM00).

Lack of a well-defined deathline is due to many uncertainties involved:

1. {\bf Criterion:} In the literatures, there exist essentially two
criteria to define the deathlines. ZHM00 adopted the criterion of $l_e 
\leq l_{acc}+l_{ph}$ (see also Zhang et al. 1997), which ensures that
the pair ``multiplicity'' is at least 1. HA01a adopted a (usually)
less demanding criterion for pair production to reach a minimum
multiplicity adequate to screen the $E_\parallel$. These could be 
regarded as the ``strong criterion''. On the other hand, HM02 
proposed a ``weak criterion''. They did not introduce any demand
on $l_e$, but just evaluated the condition that pair production
happens at all. In this criterion, $E_\parallel$ is not necessarily
screened, and the pair number density at the deathline may be very
low, i.e., could be below the GJ density. Two comments need to be
addressed. First, neither criteria are necessarily the criterion for
radio emission. The latter merely defines a pair deathline. Second,
the discrepancies between both criteria only step out in the
ICS-controlled gaps. For CR-gaps, the difference between these
treatments disappear, essentially because $l_e \ll l_{ph}$ in the CR
case (ZHM00). 

2. {\bf Model-dependence:} Clearly different models (VG or SCLF; CR-,
RICS-, or NRICS-controlled) result in different $\Delta V$'s of the
gap, and hence, different deathlines within that particular model
(e.g. ZHM00). An adequate study needs to address the parameter regimes
for the different modes to dominate so as to achieve a synthesized
deathline in the whole $P-\dot P$ space (e.g. HA01a, HM01, HM02, for
SCLF models). The following essential features are noticeable. in VGs,
usually $l_{acc}$ is small and negligible; in SCLFs, $l_{acc}$ is
comparable to (although smaller than) $l_{ph}$ for the minimized case;
in CR gaps, $l_e$ is small and negligible; in ICS gaps, $l_e$ is
comparable to $l_{ph}+l_{acc}$.

3. {\bf Equation of state (EOS):} Since 1p process is the dominant
pair formation process in normal pulsars, the magnetic fields in the
pair formation region is a crucial parameter to define the deathline
in the $P-\dot P$ space (although not in the $P-B_p$ space. For the
pure dipolar field configuration, equation (5) indicates that the
neutron star EOS influences the estimated polar cap surface magnetic
fields through influencing $I_{45}$ and $R_6$. If pair production
region is right above the surface (or a fixed altitude above the
surface), then different EOSs result in different deathlines. In SCLF
models, different EOSs also modify the frame-dragging terms. The
combined effects (HMZ02) generally makes a larger difference in short
period (e.g. millisecond) regimes than in the long period
regime. Softer neutron star EOSs or strange star EOSs tend to
facilitate pair production and hence, lower the deathlines.

4. {\bf Unknown surface field configuration:} A potentially even more
important factor that may influence the surface magnetic field
configuration and hence the deathlines is the unknown near-surface
magnetic field configuration. It has been long recognized that a
CR-controlled gap in a star-centered dipolar field can not sustain 
pair production in all known pulsars, so that near-surface multipole
magnetic fields have long been speculated (Ruderman \& Sutherland 1975;
Arons \& Scharlemann 1979). Several possible distorted magnetic
configurations have been discussed in Chen \& Ruderman (1993) within
the VG models, and these are also valid in discussing other models. 
The magnetic configuration is the key factor that influences the 
deathlines although it is sparsely modeled due to many uncertainties
involved. Recently, Gil et al. (Gil \& Mitra 2001; Gil \& Melikidze
2002; Gil et al. 2002) investigated some possible consequences when
the polar cap fields are extremely curved and strengthened.

5. {\bf Refined geometry and physics:} Most deathline discussions have 
been analytical. With the ``weak criterion'' and the assumption of a
star-centered dipole, HM01 and HM02 performed detailed numerical
simulations to study pulsar deathlines for the CR- and ICS-controlled
SCLF gaps, respectively. Refined geometry (e.g. curved spacetime) and
physics (e.g. pair formation details) are included. Their results
suggest that many of the previous analytical treatments may not be
reliable. One important finding is that at the deathlines, the maximum
usable potential is only a small fraction of the value expected
analytically (cf. ZHM00), and is model-dependent (CR vs. ICS). This
cautions us that numerical calculations may be also essential to
discuss deathlines in other models under other criteria and/or
assumptions. 

\section{A death valley in the high-$B$ regime?}

Two other QED processes can potentially suppress free pair formation 
in the strong magnetic field regime. These are magnetic photon
splitting (sp: $\gamma \rightarrow \gamma \gamma$) and bound pair
(positronium) formation. 

Baring \& Harding (1998, see also 2001, hereafter BH98, BH01)
suggested a possible radio pulsar deathline in the high magnetic field
regime, due to possible suppression of 1p pair production by magnetic
photon splitting. (Strictly speaking, pulsars do not evolve across
this line, so the ``deathline'' is essentially a radio quiescence
line.) Such a line, by definition, is defined by 
$l_{\gamma \rightarrow \gamma\gamma} < l_{\gamma\rightarrow
e^{+}e^{-}}$, where $l$'s are the $\gamma$-ray attenuation
lengths of the relevant processes (sp or 1p), which are dependent on
both the field strength and the $\gamma$-ray energy,
$\epsilon_\gamma$. By specifying a characteristic $\epsilon_\gamma$, a
deathline in the $P-B_p$ or $P-\dot P$ space can be plotted. BH98's
photon splitting deathline ($\dot P \simeq 7.9\times 10^{-13}
P^{-11/15}$) is defined by specifying $\epsilon_\gamma =
\epsilon_{esc}$, where $\epsilon_{esc}$ is critical energy of the
$\gamma$-ray that can just evade both 1p pair production and photon
splitting. 

There are several caveats concerning the BH98 photon splitting deathline:

{\bf 1. How many modes split?} The BH98 high-$B$ deathline is 
contingent upon the assumption that photons for both the $\|$ and
$\perp$ modes (defined by the electric vector with respect to the
magnetic field plain) split in superstrong magnetic fields. Within the
linear vacuum polarization treatment, only the $\perp$-mode photons
are allowed to split at least for fields below $B_q$ (e.g. Usov
2002). Whether the $\|$-mode photons split in superstrong fields due
to non-linear vacuum polarizations is fundamental for all the
discussions in this section. 

{\bf 2. High altitude pair formation?} Even if both photon modes
split, whether 1p pair production is completely suppressed still
depends on some further conditions. If particles can be accelerated at 
higher altitudes and emit photons, these $\gamma$-rays can still
produce pairs since the local field has degraded with respect to the
high value near the surface (Zhang \& Harding 2000b, hereafter ZH00b).
For a VG, on the other hand, since downwards photons can not pair
produce and there is no free source of the primary electrons,
essentially few pairs could be generated. The magnetosphere would be
dead, so strictly speaking, the photon splitting deathline is valid
only for such a case (ZH00b).

{\bf 3. Model-dependence?} Both $l_{\gamma \rightarrow
\gamma\gamma}$ and $l_{\gamma\rightarrow e^{+}e^{-}}$ depend on
$\epsilon_\gamma$ in different ways. It is more relevant to adopt the
characteristic $\epsilon_\gamma$ of a certain type of the VG gap,
rather than $\epsilon_{esc}$ to define the photon splitting deathlines
(Zhang  \& Harding 2001; Zhang 2001). The deathlines drawn with such a 
criterion tends to be tilted up with respect to the BH98 deathline in
the short period regime, since fast pulsars have larger potentials to
accelerate particles to higher energies and the more energetic photons
are more facilitated to 1p pair production (Zhang \& Harding 2001). 
Distorted magnetic configurations will also modify the competition
between the sp and the 1p processes (e.g. Gil et al. 2002).

{\bf 4. 2p pairs in magnetars?} Finally, in magnetars, even if 1p pair 
production could be completely suppressed by photon splitting, 2p pair 
production typically has a shorter attenuation length in the hot
magnetar environment (Zhang 2001). So pair production might not be
completely suppressed in magnetars. A photon splitting death valley,
if exist, is therefore only valid for the high magnetic field pulsars
without substantial magnetic decay but with strong binding at the
surface. 

Bound pair formation (e.g. Usov \& Melrose 1995; Usov 2002)
essentially delay the free pair formation front, but in principle does
not suppress free pair formation. Its role is similar to the two-mode
photon splitting processes in a SCLF gap, i.e., to increase the gap
height and the gap potential, which is helpful to interpret pulsar
high energy emission within the polar cap models.

\section{Concluding remarks}

The following statements may be pertinent:

1. We now have a clear framework about the particle acceleration and
photon-pair cascade in the pulsar polar cap region. Pair production
from the polar cap is believed to be an essential condition for
pulsar radio emission. The sufficient condition for radio emission,
however, is unknown. Personally, I think that radio emission condition 
should be more stringent than the pair production condition. Thus,
(moderate) non-dipolar fields may indeed exist at least in some
pulsars. An important advance in the pulsar study in the recent years
is the realization that ICS plays a crucial role in stead of CR in
determining gap properties at least in some pulsars. 

2. Theories are generally successful to define the conventional radio
pulsar death in the long period regime, although many uncertainties
prevent us from achieving a solid deathline. A death valley is more
pertinent. Without introducing distortions from a star-centered-dipolar 
configuration, one can not include all pulsars above the deathline. 
The 8.5 second pulsar (Young et al. 1999) remains a challenge for any
pure-dipole model after detailed numerical simulations. A
best guess is that an ICS-controlled gap anchors in this pulsar with
a moderate non-dipolar near-surface field configuration.

3. In the high magnetic field regime, pulsar death is not
unambiguously defined. There is no strong reason against the possible
radio emission from high magnetic field pulsars and magnetars. 
Possible reasons of apparent radio quiescence of magnetars may be due
either to (a) that the coherent condition is destroyed; or to (b) that
the main energy band of the coherent emission is not in radio; or to
(c) the beaming effect; or else to (d) that the soft gamma-ray
repeaters and the anomalous X-ray pulsars are not magnetars at all
(but might be accretion-powered systems).

\acknowledgements {I am grateful to Alice Harding, G. J. Qiao, and Alex
Muslimov for stimulating collaborations, to Jon Arons, Matthew Baring,
Janusz Gil for discussions concerning various topics covered in this
talk, and to R. X. Xu for presenting the materials at the meeting on
behalf of me.}

\end{document}